# Field-controlled dynamics of skyrmions and monopoles

Jung-Shen B. Tai[1], Andrew J. Hess[1], Jin-Sheng Wu[1] and Ivan I. Smalyukh[1,2,3*]

[1]*Department of Physics and Chemical Physics Program, University of Colorado, Boulder, CO 80309, USA*

[2]*Department of Electrical, Computer, and Energy Engineering, Materials Science and Engineering Program and Soft Materials Research Center, University of Colorado, Boulder, CO 80309, USA*

[3]*Renewable and Sustainable Energy Institute, National Renewable Energy Laboratory and University of Colorado, Boulder, CO 80309, USA*

*\* Correspondence to:  ivan.smalyukh@colorado.edu*

**Magnetic monopoles, despite their ongoing experimental search as elementary particles, have inspired the discovery of analogous excitations in condensed matter systems. In chiral condensed matter systems, emergent monopoles are responsible for the onset of transitions between topologically distinct states and phases, like in the case of transitions from helical and conical phase to A-phase comprising periodic arrays of skyrmions. By combining numerical modeling and optical characterizations, we describe how different geometrical configurations of skyrmions terminating at monopoles can be realized in liquid crystals and liquid crystal ferromagnets. We demonstrate how such complex structures can be effectively manipulated by external magnetic and electric fields. Furthermore, we discuss how our findings may hint at similar dynamics in other physical systems, and their potential applications.**

**Introduction**

Magnetic monopoles, hypothetical elementary particles, which are the point sources of magnetic fields, are the holy grail of particle physics. Despite promising restoration of duality symmetry in the electromagnetic theory and various theoretical predictions, the experimental discovery of magnetic monopoles has remained elusive(*1, 2*). Like in the case of macroscopic magnets, the sources and sinks of magnetic field could not be found separated from each other at the level of elementary particles. However, structures analogous to monopole-like excitations have been recently discovered in various condensed matter systems, including spin ices(*3*), Bose-Einstein condensates(*4*) and magnetic solids(*5*). Such excitations are particularly useful in understanding emergent behavior of these condensed matter systems and physically resemble the elementary-particle counterparts of magnetic monopoles beyond topological classification (as a matter of fact, monopoles as topological objects are rather common). In magnets, $\pi_2(\mathbb{S}^2)$ topological point defects are monopoles not just in the magnetization field $\boldsymbol{m}(\boldsymbol{r})$, but also in the emergent magnetic field, $(B_{\text{em}})_i \equiv \hbar \varepsilon^{ijk} \boldsymbol{m} \cdot (\partial_j \boldsymbol{m} \times \partial_k \boldsymbol{m})$, a fictitious field describing the interaction between conduction electrons and the underlying spin texture(*6–8*). Emergent monopoles in chiral ferromagnets play an important role during a phase transition between topologically distinct phases, serving as nucleation points for the inception of topological transformation(*5*). In liquid crystals (LCs) and LC ferromagnets, two- and three-dimensional topological point defects are ubiquitous(*9, 10*). Topological defects in LCs also serve as a means to embed topological solitons of lower dimensions in the topologically trivial background of a higher dimensional space(*11*). A few examples include embedding $\pi_2(\mathbb{S}^2)$ skyrmions in the three-dimensional space as torons and $\pi_1(\mathbb{S}^1)$ twist walls in the two-dimensional space as one of the common cholesteric finger structures and recently observed configurations of Möbiusons(*11, 12*).

In this article, using externally applied magnetic and electric fields, we demonstrate the dynamical control of $\pi_2(\mathbb{S}^2)$ skyrmions and the accompanying $\pi_2(\mathbb{S}^2)$ topological defects, monopoles in the emergent field. We show that, much like the skyrmion A-phase, periodic lattices of skyrmions can also form in LCs when embedded within either a uniform or a helical background. Moreover, we find that skyrmions and monopoles can be effectively controlled by external fields while the motions of the monopoles are directly observed with an optical microscope with which observations benefit particularly from in-plane configurations of skyrmions. Furthermore, we demonstrate that in a chiral LC material, the vertical skyrmion terminating at monopoles (the so-called torons) can be hosted in a helical or uniform field background with controllable inter-soliton interaction, modulated by the applied voltage. While the analogy of our skyrmions and monopoles to their high-energy-physics counterparts like Dirac monopoles(*13*) and $\pi_3(\mathbb{S}^3)$ Skyrme solitons(*2*), which in particle-physics models describe subatomic particles with different baryon numbers, is distant, the observed findings closely mimic what was found in chiral magnets (*5*) where even its Hamiltonian can have a form similar to that of a chiral LC, so that LCs could potentially serve as model systems and provide useful insights. The diversity of observed geometrically distinct but topologically equivalent torons, or monopole-terminating skyrmions, points to possible topology-preserving switching between them, in analogy to the recently studied inter-transformation between geometrically distinct states of hopfions(*14*). Finally, we discuss the fundamental implications of monopole control in LC systems and their potential technological applications.

**Results**

**$\pi_2(\mathbb{S}^2)$ point defects are monopoles in emergent field that embed $\pi_2(\mathbb{S}^2)$ skyrmions in**

**topologically trivial backgrounds**

Ubiquitously observed in the order-parameter field $\boldsymbol{n}(\boldsymbol{r})$ of condensed matter systems, a hedgehog point defect and a hyperbolic point defect are both topological point defects classified by the $\pi_2(\mathbb{S}^2)$ homotopy group and have a topological charge of unity (Fig. 1, A and B). In the emergent magnetic field, despite the distinct difference in their geometrical field configurations, both hedgehog and hyperbolic defects are monopoles – radial sources of emergent magnetic fields (Fig. 1, A-C). In LCs, LC ferromagnets, and solid-state magnets, torons are $\pi_2(\mathbb{S}^2)$ skyrmions terminating at hyperbolic point defects of the same $\pi_2(\mathbb{S}^2)$ field topology, often observed in a uniform field background with a perpendicular boundary condition (Fig. 1D)(*10, 15–17*). Note that since all $\boldsymbol{n}(\boldsymbol{r})$ structures discussed in this work can be consistently vectorized even when the order parameter of the system has nonpolar head-tail symmetry, namely $\boldsymbol{n}(\boldsymbol{r}) \equiv -\boldsymbol{n}(\boldsymbol{r})$ and the corresponding order-parameter space is $\pi_2(\mathbb{S}^2/\mathbb{Z}_2)$, we focus on vectorized $\boldsymbol{n}(\boldsymbol{r})$ exclusively for simplicity(*18*). The two chiral hyperbolic point defects – deformable from the hyperbolic defect in Fig. 1B (or its oppositely-charged counterpart by reversing $\boldsymbol{n}(\boldsymbol{r})$ direction) and is thus topologically equivalent – act as a topological switch that transitions between regions of trivial ($N_{sk} = 0$) and non-trivial topology ($|N_{sk}| = 1$) and are required to embed skyrmions in the trivial uniform background (Fig. 1, D and E). Here $N_{sk}$ is the skyrmion number of the 2D $\boldsymbol{n}(\boldsymbol{r})$ field with the definition, $N_{sk} \equiv \frac{1}{4\pi} \int d^2\boldsymbol{r}\, \boldsymbol{n} \cdot (\partial_x \boldsymbol{n} \times \partial_y \boldsymbol{n})$. The emergent field $\boldsymbol{B}_{em}$ derived from $\boldsymbol{n}(\boldsymbol{r})$ of a toron shows the corresponding two defects in $\boldsymbol{B}_{em}$ are a pair of a monopole and an anti-monopole, acting as the source and sink of the emergent field (Fig. 1F). The $\boldsymbol{B}_{em}$ flux streaming from the monopole to the anti-monopole is reminiscent of Dirac's string in Dirac monopole(*13*). Therefore, the emergence of a skyrmion with non-trivial $\pi_2(\mathbb{S}^2)$ field topology in a trivial field background is mediated by the $\pi_2(\mathbb{S}^2)$ defects, which are a pair of monopoles in $\boldsymbol{B}_{em}$.

**In-plane skyrmions and monopoles can be stabilized in a uniform or helical background**

In chiral materials such as chiral LCs and chiral magnets, skyrmions are often recognized as axisymmetric whirling configurations that smoothly transition into a uniform far-field $n_0$. Such uniform-far-field skyrmions can be approximated analytically by the hedgehog ansatz $n(r) = (\sin f(r)\cos(\theta+\gamma), \sin f(r)\sin(\theta+\gamma), \cos f(r))$, where $f(r)$ is a monotonic function with $f(0) = \pi$ and $f(\infty) = 0$, $\theta$ is the polar angle of the 2D spatial coordinate $r$, and $\gamma = \pi/2$ for right-handed chiral material. The corresponding $B_{\text{em}}$ (or equivalently, skyrmion number density) derived from the ansatz also possesses the axial symmetry (Fig. 2, A and C). Such skyrmions in a uniform background are often found in the horizontal mid-plane perpendicular to $n_0$ in chiral LCs and chiral LC ferromagnets between substrates with a perpendicular (homeotropic) boundary condition, where torons emerge as a result of geometrical frustration between the material's chirality and the substrate's surface alignment (Fig. 1, D-F)(*11*). A somewhat less commonly familiar configuration of skyrmions is that with a helical field background; in this case, the corresponding $n(r)$ can be thought of as splitting and splicing the uniform-background skyrmion while extending the region of helical configuration into a uniformly helical background with constant helical axis $\chi_0$, which defines the axis that $n(r)$ twists around in the far field (Fig. 2, B and D)(*19*). Helical skyrmions constructed as such show distinct symmetry in the distribution of skyrmion number density where it is localized at two lobes around the so-called $\lambda$ disclination lines (often also referred to as fractional skyrmions and merons), the singular line defects in the helical-axis field that remain nonsingular in $n(r)$ (*20*). Note in the case of helical skyrmion configurations, which in LC literature are sometimes referred to as Lehman clusters, the helical background is also topologically trivial and allows for compactification of the configuration space $\mathbb{R}^2$ to $\mathbb{S}^2$ and thus the classification of the observed configuration within $\pi_2(\mathbb{S}^2)$ topology.

The richness in geometrical configurations of skyrmions expands the confinement and alignment conditions that can host the solitons and associated monopoles. In particular, torons and skyrmions can also emerge in cells with planar alignment, where their stability could be established by frustration induced by cholesteric compression/dilation or by coupling to externally applied electric and magnetic fields. Computer simulations based on Frank-Oseen free-energy minimization show skyrmions stabilized both in a uniform-field background and in a helical-field background, respectively (Fig. 2, E and H). The $\pi_2(\mathbb{S}^2)$ topological defects in $\boldsymbol{n}(\boldsymbol{r})$ mediate the embedding of skyrmions into the field backgrounds (Fig. 2, F and I), and are sources and sinks of $\boldsymbol{B}_{\text{em}}$ where the streamlines of $\boldsymbol{B}_{\text{em}}$ run along the in-plane direction parallel to the substrates, representing monopoles of opposite charges (Fig. 2, G and J). Despite the dissimilar $\boldsymbol{n}(\boldsymbol{r})$ configurations around the defects, interestingly, they are classified by the $\pi_2(\mathbb{S}^2)$ homotopy group and correspond to consistent charges of monopoles in $\boldsymbol{B}_{\text{em}}$. Compared to the more commonly realized skyrmions in perpendicular-alignment cells, in-plane skyrmions and monopoles allow the same topology to be realized in a geometry where they could reconfigure in the direction parallel to the confining substrates; this allows better accessibility for optical characterization and freedom of dynamics for monopoles.

**In-plane skyrmion emergence under applied electric field and their 1D lattices**

To observe in-plane skyrmions experimentally, we confined chiral LCs between substrates with planar anchoring in LC cells with a thickness-to-pitch ratio of $d/p = 1$. Under such conditions, the uniformly helical (cholesteric) state with far-field helical axis $\boldsymbol{\chi}_0$ perpendicular to the substrates is the global ground state and no stable skyrmions or monopoles are observed. Upon heating and quenching from the isotropic state to the cholesteric state, dark fragments of stripe-like patterns seen under a polarizing optical microscope (POM) between crossed polarizers can be stabilized when a voltage across the substrates of $U \approx 1.4$ V is applied to the LC with positive dielectric anisotropy (Fig. 3A). To reveal unambiguously the $\mathbf{n}(\mathbf{r})$ of these fragments, we employed a nonlinear fluorescence polarizing imaging (3PEF-PM, *Methods*) where the generated patterns can be compared with those from $\mathbf{n}(\mathbf{r})$ simulations. The results obtained using different polarizations of excitation are consistent with $\mathbf{n}(\mathbf{r})$ of in-plane skyrmions terminating at monopoles in a helical background (Fig. 3B). The in-plane helical skyrmions are stabilized by an electric field $\mathbf{E}$ because of the skyrmion's energetically favored vertical $\mathbf{n}(\mathbf{r})$ along $\mathbf{E}$, which is not present in the uniformly helical background (Fig. 2H).

Besides spontaneous emergence after quenching, in-plane skyrmions can also nucleate from sample edges or be controllably generated by local melting and reorienting of LCs using laser tweezers when an electric field is applied (*Methods*). As the skyrmions increase in number and fill the sample, they form a 1D lattice in the direction perpendicular to the direction of planar alignment (Fig. 3, C and D). Defects of the skyrmion lattice as quasi-dislocations can be observed readily in a large lattice, where skyrmions terminating at monopoles locally disrupt the periodic arrangement (Fig. 3E). 3PEF-PM images show changes in $N_{\text{sk}}$ in the vertical cross-sections mediated by monopoles, where the presence of a monopole is associated with a change in $N_{\text{sk}}$ by unity. Specifically, Fig. 3G shows the effect of topological switching mediated by monopoles: $N_{\text{sk}}$ increases from -3 to -2 by a monopole, and decreases from -2 to -3 by an anti-monopole (Fig. 3, E-G). The 1D skyrmion lattice in chiral LCs is analogous to the 2D skyrmion A-phase in chiral magnets, where both form as a result of energy competition between material's free energy minimization and coupling to externally applied fields. Notably, the unwinding of the chiral magnetic A-phase into ferromagnetic phase is also mediated by monopoles(*5*). The energetics of the emergence of skyrmions/torons mediated by monopole-antimonopole nucleations in this geometry competes with that of the formation of undulations of chiral LC's quasilayers, where they, for certain material parameters in a similar experimental geometry, can emerge as a transformation that preserves the trivial topology of the helical background (*21*).

In-plane skyrmions can also be realized in a uniform-field background (Figs. 2E and 3H,I). For a confined chiral LC with planar anchoring and $d/p \approx 1$, an in-plane electric field ~1 V/μm unwinds the helical $n(r)$ configuration and makes the uniform state $n_0 \parallel E$ the ground state. Uniform-background in-plane skyrmions and monopoles can thus be created spontaneously through quenching or controllably by laser tweezers. Compared to the case of helical-background skyrmions, the stabilization of uniform-background skyrmions with an electric field requires high voltages (at least ~ 1000 V) because of the large lateral dimensions compared to the thickness in our LC samples. However, more sophisticated sample geometry such as those with patterned electrodes that reduce the gap size should alleviate the need for high voltages.

**External fields control dynamics of skyrmions and monopoles in LCs and LC ferromagnets**

Stabilization of in-plane skyrmions by an electric field $E$ is enabled by the balance between elastic energy and dielectric coupling energy, where dielectric coupling energetically favors $n(r) \parallel E$ in a material with positive dielectric anisotropy. In the case of helical-background skyrmions, $E \parallel \hat{z}$ promotes vertical $n(r)$ orientations and the expansion of skyrmions, which counters the shrinking tendency due to mismatch between twist deformation in the skyrmion and molecular pitch $p$. Beyond simply stabilizing $n(r)$ configurations, modulation of $E$ can also be used to control the dynamics of monopoles on which the skyrmions terminate. Here we focus on in-plane helical-background skyrmions because of their ease of manipulation and observation, though the same field-controlled dynamics can be straightforwardly applied to other geometrical configurations.

For a pair of a monopole and an anti-monopole at the ends of a skyrmion fragment, we define the separation velocity as the rate in the change of their distance of separation. Away from the equilibrium voltage at $U < 1.4$ V, the dielectric coupling is not strong enough to compensate for the cost in elastic energy in the skyrmion, and the skyrmion shrinks (Fig. 4A and movie S1). Whereas at $U > 1.4$ V, it becomes favorable in the overall free energy for the skyrmion to grow and acquire larger regions of vertical orientations in the system (Fig. 4B and movie S1). Thus, the separation velocity becomes positive and the width of the skyrmion, as determined in POM by the region deviating away from the helical background, is also visibly larger. The dependence of separation velocity on $U$ is plotted in Fig. 4C. The magnitude of this separation velocity $\approx -60$ µm/s is largest when the skyrmion shrinks at $U = 0$ V, while at $U = 1.6$ V, the monopole pair separate at 20 µm/s. In the range $U < 1.6$ V, the skyrmions connecting monopole pairs remain approximately linear in shape. Above $U = 1.6$ V, distortion in the helical background takes place in the form of undulations(*21*), resulting in deformation of the linear shape of the skyrmion and obstruction in the separation of monopoles. We note that when $U < 1.4$ V, the skyrmion shrinks and the monopole pair come together and annihilate, indicating the absence of an energy barrier between the monopole pairs. This contrasts with topological solitons, where $\boldsymbol{n}(\boldsymbol{r})$ is continuous and topologically nontrivial; erasing them requires generation and annihilation of singular defects, leading to an energy barrier and the so-called topological protection(*20*).

The dynamics and stability of skyrmions and monopoles can also be controlled by magnetic fields in chiral LC ferromagnets where magnetically monodomain nanoplates are uniformly dispersed at high concentrations in the hosting chiral LC (*22*). By forming a monodomain in the magnetization $m(r)$ of ferromagnetic nanoplates and introducing strong homeotropic surface anchoring of LC molecules to the nanoplates (*Methods*), $m(r)$ follows $n(r)$ and physically vectorizes $n(r)$ from a director field into a vector field. When an external magnetic field $H$ is applied, $H$ and $m(r)$ couple linearly and compete with free-energy contributions from elastic and dielectric interactions. Figure 4, D and E, show the magnetic control of skyrmions and monopoles with $H$ applied in different directions. When $H$ is parallel to the average $m(r)$ within a skyrmion, $m_s$, which is also the magnetization at the top and bottom substrates (Fig. 2H), the linear magnetic coupling promotes the expansion of skyrmions and separation of monopoles (Fig. 4D). The width of the skyrmions also become visibly wider. When $H$ is antiparallel to $m_s$, skyrmions contract and eventually monopole pairs annihilate (Fig. 4E). Compared to electric-field control of monopoles, the coupling between $H$ and $m(r)$ is linear and polar, and can be applied with less constraint pertaining to sample geometry, enabling an orthogonal dimension in controlling monopole dynamics.

**Electric field controls interaction and assembly of vertical torons by modulating energetic landscape around solitons**

Thus far, we have focused on in-plane skyrmions and monopoles because of their accessibility for manipulation and observation. However, vertical skyrmions along the normal of substrates and terminating at monopoles in a cell with homeotropic boundary conditions, also known as torons, also display manipulable interaction and assembly modulated by an electric field. In a homeotropic cell, when the confined chiral LC has an elastic anisotropy such that $K_{33}/K_{22} \leq 1$, i.e., the energetic penalty of bend deformation is smaller than that of twist deformation, the background $n(r)$ can be switched between the helical state and the uniform state by an electric field. For a material with positive dielectric anisotropy, the background is helical at no field and can be unwound to the uniform state when an electric field is applied (*14*, *23*). Vertical skyrmions stabilized in a helical background with no field display attractive pair-wise interaction and form a close-packed hexagonal lattice when several of them are in close proximity (Fig. 5A). This is in stark contrast to the repulsive interaction between vertical skyrmions stabilized in a uniform background(*15*, *18*). When an electric field across the thickness of an LC cell was applied to unwind the helical background into a uniform background, the skyrmion-skyrmion interaction switched from attractive to repulsive (Fig. 5B). Computer simulations based on free-energy minimization reveal the detailed $\boldsymbol{n(r)}$ transformation when the background state changes between the helical state and the uniform state (Fig. 5, C and D). Before and after the transformation of the background state, the monopole pairs and skyrmion number in the horizontal cross-sections are conserved, preserving their respective $\pi_2(\mathbb{S}^2)$ topology. However, the energetic landscape in the lateral dimensions changes. When the background is helical at no electric field, the free energy density at the periphery of the skyrmion is higher than that of the background; when the background is unwound by an electric field, the free energy density at the periphery of the skyrmion becomes lower than that of the background (Fig. 5E). To minimize total free energy, skyrmions (in a helical background) with a high-energy periphery comes closer together to share and reduce the volume with a high energy density, leading to an attractive interaction, whereas skyrmions (in the uniform background) with a low-energy periphery stay away from each other to increase the space with low energy density, leading to repulsion. We note that, in noncentrosymmetric chiral magnets, skyrmions in a uniformly polarized background are commonly observed, and skyrmions with attractive interaction have also been demonstrated in a conical background(*24*). In LCs, the switching of pair-wise interactions between skyrmions through background modulation can be recorded in real time due to their unparalleled experimental accessibility.

**Discussion and Conclusion**

In this work, we have presented the magnetic and electric control of the equilibrium configuration of $\pi_2(\mathbb{S}^2)$ topological configurations, namely skyrmions and monopoles (with their composite structures referred to as torons), and their dynamics in chiral LC materials. We found that $\pi_2(\mathbb{S}^2)$ point defects at the termini of skyrmions are indeed monopoles in the emergent magnetic field. We further demonstrated that LC skyrmions can take on versatile geometrical configurations under different boundary conditions and applied fields. Notably, in-plane skyrmions in planar-confined cells offer exceptional accessibility for observation and manipulation by external fields. Compared to vertical skyrmions in a homeotropically aligned cell (torons), in-plane skyrmions offer the opportunity of controlling the separation of monopoles in the lateral dimensions, whereas the separation between the monopoles in vertical skyrmions is limited due to confinement by the substrates, though their lateral spatial extent can be controlled by fields(*10*, *25*). To enhance the dynamical range of the electric control of skyrmions and monopoles, dual frequency LCs may be used to allow the sign of LC dielectric anisotropy to be adjustable, such that positive (negative) dielectric anisotropy that promotes expansion (contraction) can be used in the same material(*26*). The linear coupling enabled by magnetic fields, distinct from the quadratic nature of dielectric coupling to electric fields, could also potentially allow for orientational control of solitons and monopoles.

Additionally, we showed that the pair-wise interaction between vertical skyrmions can be switched between attractive and repulsive, by altering the background configuration that embeds them through applying electric fields. We expect that similar switching can be achieved as well in LC ferromagnets using magnetic fields, where the far-field $m(r)$ can be polarized by $H$ applied to the sample. Crossover between different pair-wise interactions of LC skyrmions has been observed between vertical and in-pane skyrmions in homeotropically aligned cells, albeit it was done differently through changing the dimension of the confined LC or anchoring(*27*). The field-controlled pair-wise interaction of LC skyrmions could potentially further couple with their recently observed squirming motion, activated by a modulated electric field, to facilitate versatile control and manipulation of LC topological solitons and defects(*28*).

On the application side, the high degree of distortion in the alignment field around the monopole and the central isotropic core are elastic energy traps for micro- or nanoparticles(*29*, *30*). Thus, potential technological applications could arise from dynamical control of skyrmions and monopoles, such as electro/magneto-optic devices, microfluidic devices, optical logic devices, etc. The widespread electro-optic applications of LCs typically rely on switching between topologically trivial states that are homeomorphic one to another, but our findings point to the possibility of realizing controlled conditions for achieving novel electro-optic effects either by morphing topologically nontrivial configurations or even by changing topology, for example, through the monopole-mediated skyrmion nucleation and reconfiguration. Lastly, the study of dynamics and control of LC skyrmions and monopoles could shed light on the physics of topologically equivalent solitonic structures in other ordered systems, such as in the solid-state magnetic(*5*), ferroelectric(*31*), or optical fields(*32*).

**Figures**

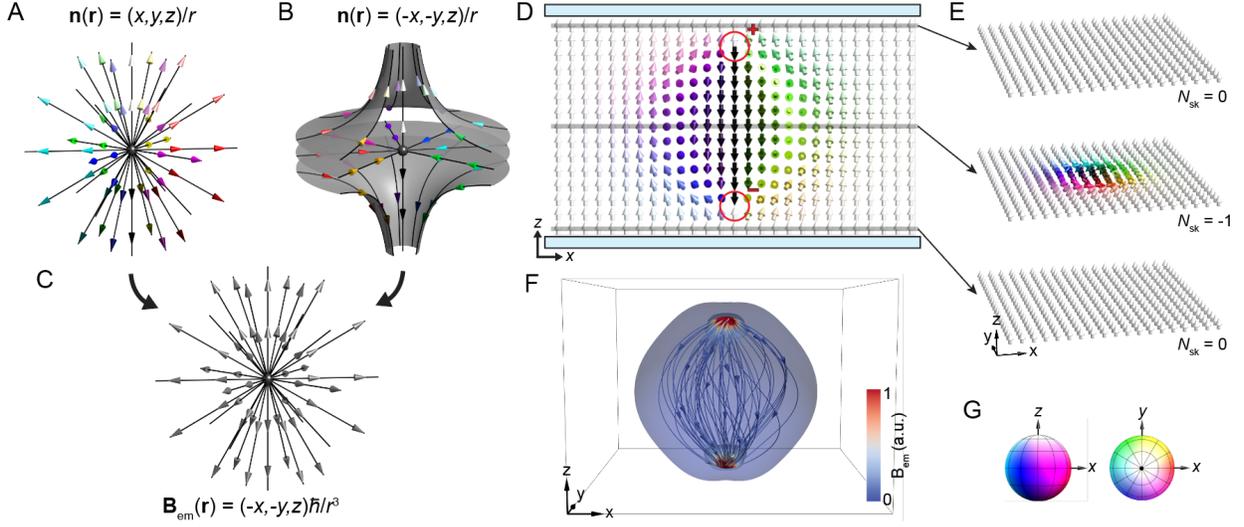

**Fig. 1. Monopoles and point defects.** (**A**,**B**) A hedgehog (A) and a hyperbolic (B) point defect in the material alignment field $\boldsymbol{n}(\boldsymbol{r})$. (**C**) Emergent magnetic field $\boldsymbol{B}_{em}$ derived from the ansatzes of the hedgehog (A) and the hyperbolic point defect (B) in $\boldsymbol{n}(\boldsymbol{r})$ both show radial monopole-like field configuration. (**D**,**E**) Computer-simulated $\boldsymbol{n}(\boldsymbol{r})$ configuration of a toron in a cell with a perpendicular boundary condition shown in the vertical cross-section parallel to the uniform far-field $\boldsymbol{n}_0 \parallel \hat{z}$ (D) and horizontal cross-sections at different positions along $z$ indicated in (D) and their respective skyrmion number $N_{sk}$. Point defects (monopoles) in (D) are circled in red with their charges labeled. Here $\boldsymbol{n}(\boldsymbol{r})$ is visualized by colored arrows using the color scheme in (G). The blue slabs indicate the substrates confining LCs. (**F**) $\boldsymbol{B}_{em}$ or equivalently the 3D skyrmion number density derived from the $\boldsymbol{n}(\boldsymbol{r})$ configuration of the toron in (D) shown by the isosurfaces of the scalar magnitude of $\boldsymbol{B}_{em}$ and streamlines of $\boldsymbol{B}_{em}$ connecting the positively-charged (top) and the negatively-charged (bottom) monopoles. (**G**) Order-parameter space $\mathbb{S}^2$ colored according to the vectorized $\boldsymbol{n}(\boldsymbol{r})$ orientation.

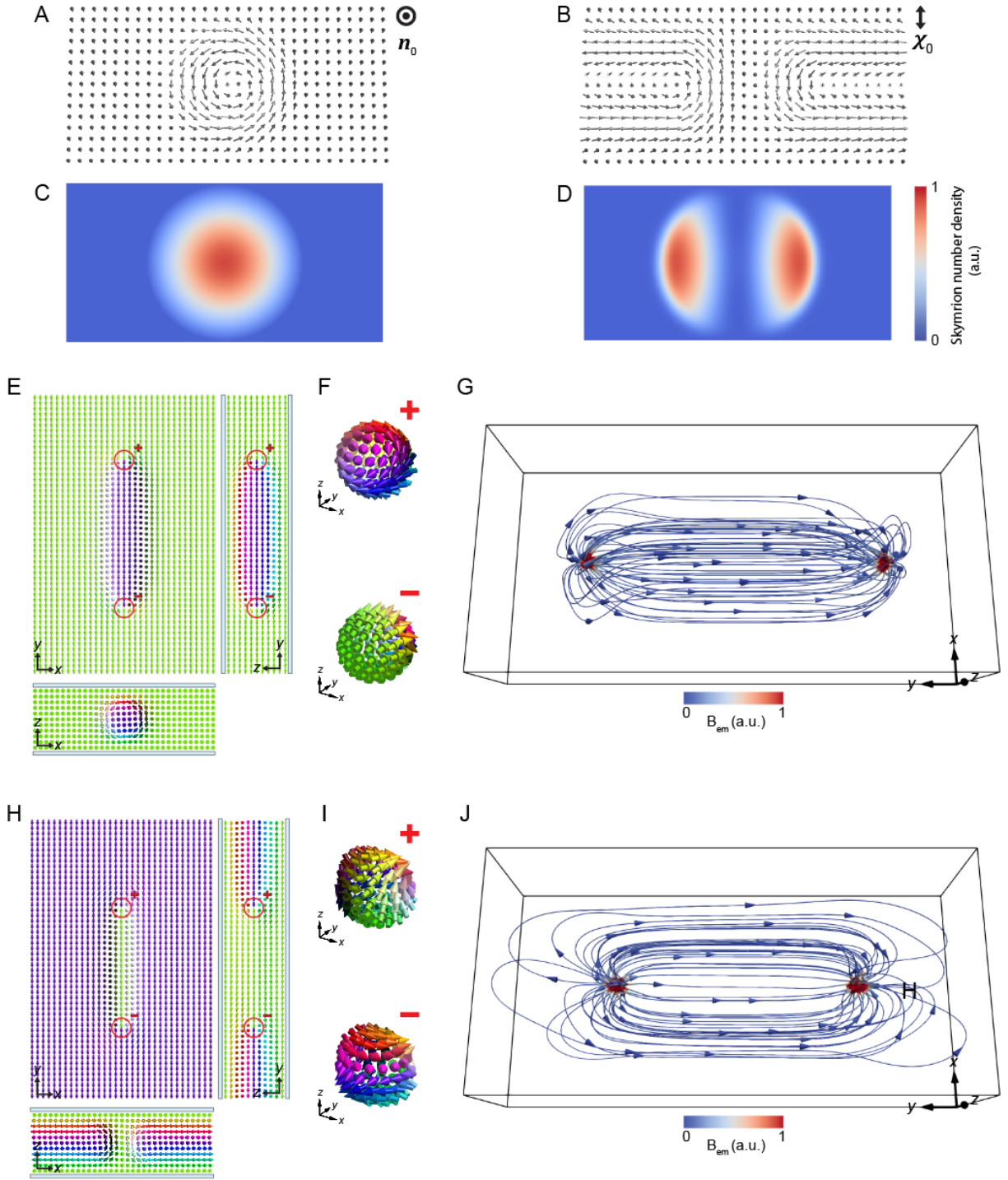

**Fig. 2. Toron in a uniform or helical background and the associated monopoles in emergent magnetic field. (A,B)** Torons are formed by monopole-associated skyrmions which can be embedded in a uniform far-field $\mathbf{n}_0$ (A) and a helical field background with constant helical axis $\boldsymbol{\chi}_0$ (B), respectively. **(C,D)** Skyrmion number density derived from $\mathbf{n}(\mathbf{r})$ in (A) and (B), respectively. **(E)** Computer-simulated cross-sections of $\mathbf{n}(\mathbf{r})$ of an in-plane skyrmion terminating at two monopoles (circled in red with the sign of the charge labeled) and embedded in a uniform

far-field $\boldsymbol{n}_0 \parallel \hat{y}$. (**F**) Detailed $\boldsymbol{n}(\boldsymbol{r})$ around each monopole in (E). (**G**) Emergent field $\boldsymbol{B}_{em}$ derived from $\boldsymbol{n}(\boldsymbol{r})$ in (E) shown by streamlines of $\boldsymbol{B}_{em}$ colored by its magnitude. (**H**) Computer-simulated cross-sections of $\boldsymbol{n}(\boldsymbol{r})$ of a skyrmion terminating at two monopoles (circled in red with the sign of the charge labeled) and embedded in a helical far-field with helical axis $\boldsymbol{\chi}_0 \parallel \hat{z}$. (**I**) Detailed $\boldsymbol{n}(\boldsymbol{r})$ around each monopole in (H). (**J**) Emergent field $\boldsymbol{B}_{em}$ derived from $\boldsymbol{n}(\boldsymbol{r})$ in (H) shown by streamlines of $\boldsymbol{B}_{em}$, colored by its magnitude. In simulations, $d/p = 0.8$ in (E-G), and $d/p = 1$ and $U = 1.5$ V in (H-J), respectively.

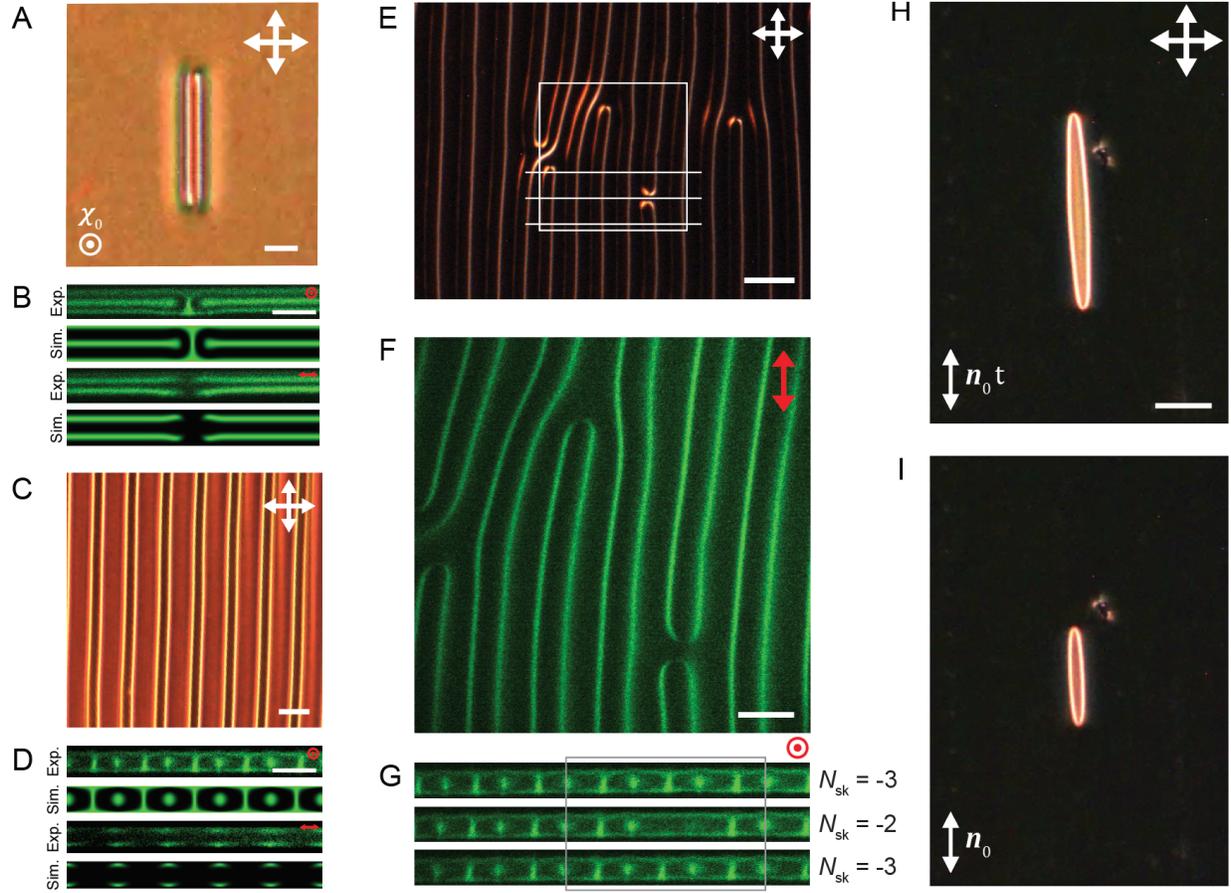

**Fig. 3. Torons comprising skyrmion and monopoles in confined chiral LCs.** (**A**) POM images of an individual toron with a skyrmion (A) in a helical far-field background (constant helical axis $\chi_0$) terminating at two monopoles (**B**) Experimental (exp.) and computer-simulated (sim.) 3PEF-PM images in the vertical cross-section through the skyrmion. (**C**) 1D lattice of skyrmions in a helical background. (**D**) Experimental (exp.) and computer-simulated (sim.) 3PEF-PM images in the vertical cross-section through an 1D lattice of helical-background skyrmions. (**E**) POM image of a skyrmion lattice with defects. The sample in (E) was polymerized and washed before imaging, reducing the birefringence. (**F,G**) 3PEF-PM images of skyrmions and monopoles in horizontal and vertical cross-sections indicated in (E). The skyrmion number $N_{sk}$ of each vertical cross-section in the region defined by the gray box are shown. (**H,I**) POM images of an individual skyrmion in a uniform far-field background $n_0$ terminating at two monopoles with variable length under a smaller (H) or a larger (I) in-plane electric field. Scale bars are 10 μm in A, B, D, F and 30 μm in C, E, H. Sample thicknesses are 30 μm in (C) and (E) and 10 μm otherwise. $d/p = 1$ in all samples. The polarization states of the linearly polarized excitation light for 3PEF-PM in B,D,F,G are marked in red.

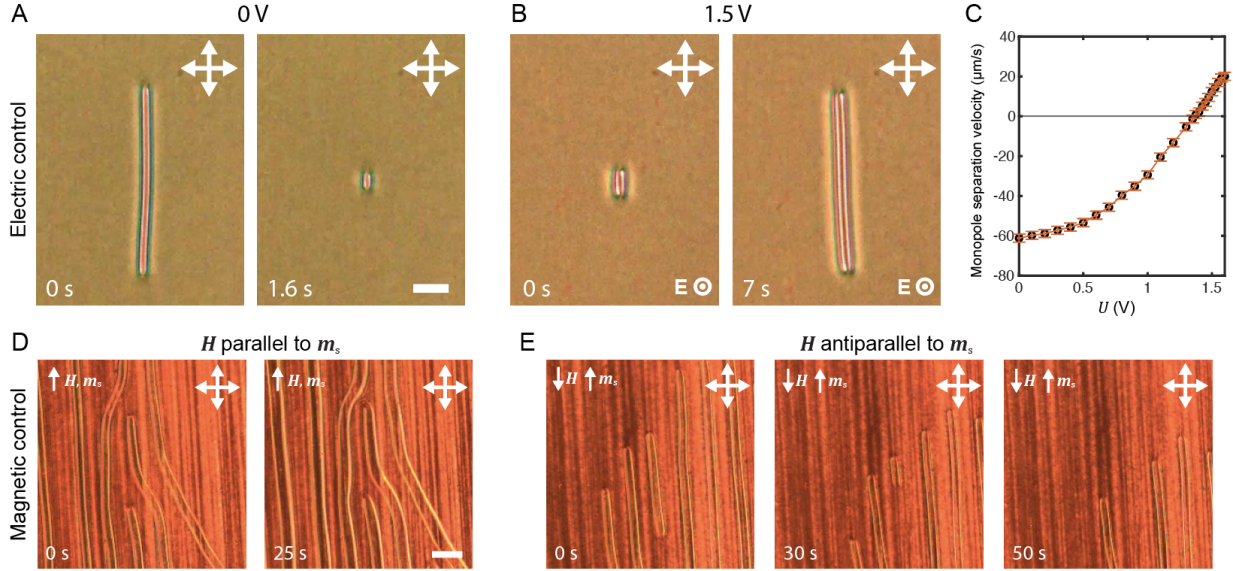

**Fig. 4. Field-controlled dynamics of monopoles at skyrmion ends.** (**A,B**) Snapshots of POM images of two monopoles connected by a skyrmion in a helical background moving close to (A) and away from (B) each other at $U = 0$ V and $U = 1.5$ V, respectively. (**C**) Separation velocity of the monopole-antimonopole pair vs. the applied voltage $U$. The error in measuring velocity is $\pm 2$ μm/s. (**D,E**) Snapshots of POM images of two monopoles connected by a skyrmion in a helical background moving close to (D) and away from (E) each other when an external magnetic field $\boldsymbol{H}$ is applied parallel or antiparallel to magnetization $\boldsymbol{m}_s$, the average $\boldsymbol{m}(\boldsymbol{r})$ within the skyrmion, respectively. Scale bars are 20 μm in (A,B) and 50 μm in (D,E). Sample thicknesses are 10 μm in (A) and (B) and 60 μm in (D) and (E). $d/p = 1$ in all samples.

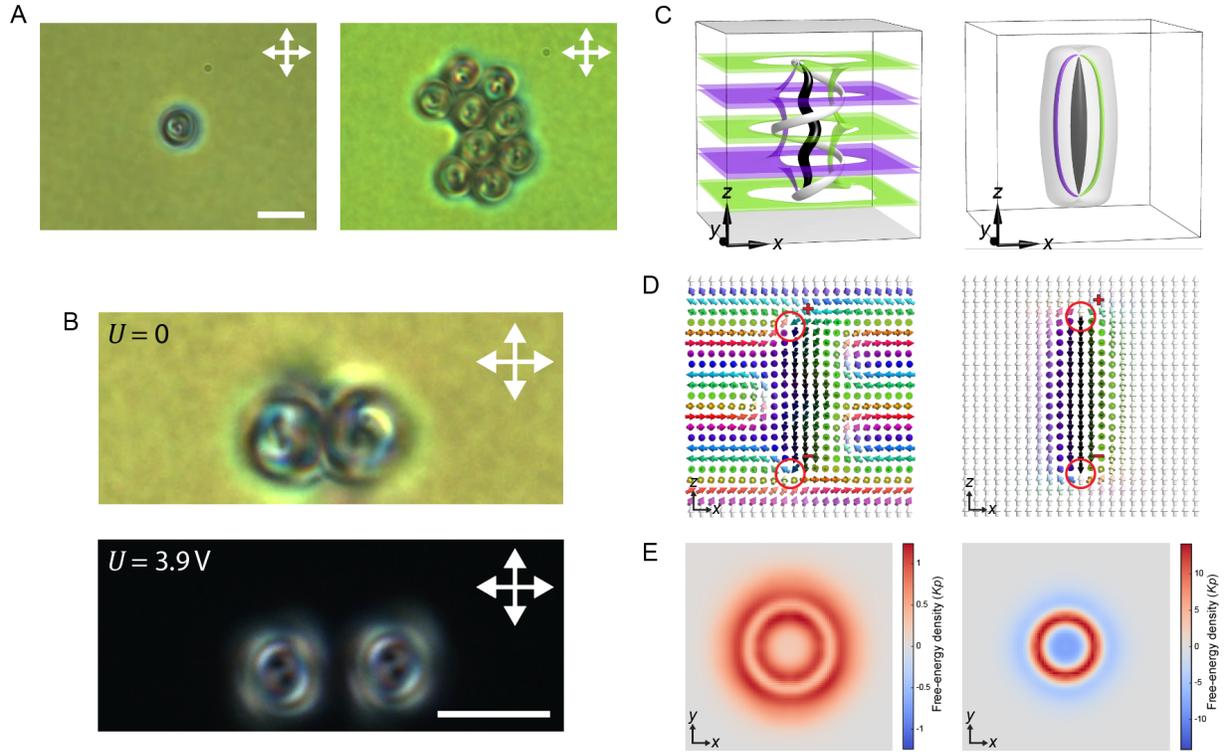

**Fig. 5 | Controlling pair-wise interaction between vertical skyrmions.** (**A**) POM images of a vertical skyrmion in a helical background with perpendicular boundary conditions individually (left) and aggregating into a close-packed crystallite (right). (**B**) The pair-wise interaction between skyrmions is altered by changing the embedding far-field between helical (up) and uniform (bottom) through modulating the applied voltage $U$. (**C**) Preimages of $\boldsymbol{n}(\boldsymbol{r})$ orientations along $\pm\hat{z}$ and $\pm\hat{y}$ colored according to order-parameter space $\mathbb{S}^2$ shown in Fig. 1G for computer-simulated skyrmions in a helical background (left) and a uniform background (right), respectively. (**D**) Computer-simulated vertical cross-section of $\boldsymbol{n}(\boldsymbol{r})$ of a vertical skyrmion terminating at monopoles in a helical (left) and a uniform (right) background, respectively. Monopoles are circled in red with their charges labeled. (**E**) Free-energy density of the computer-simulated skyrmion field configurations shown in (C and D) in a helical (left) and a uniform (right) background, respectively. The energy density is averaged over thickness ($z$ direction) and excludes regions containing singular monopoles, shown in units of $Kp$, the product of the average elastic constant and pitch. In experiments and simulations, $d/p = 3$ and $d = 10$ μm. $\sqrt{K_{33}/K_{22}} = 0.8$ and $U = 4.1$ V were used in simulations. Scale bars are 5 μm.

**Methods**

**Sample preparation**

Chiral LC mixtures used in this work comprise 4-Cyano-4'-pentylbiphenyl (5CB, from EM Chemicals) added with the right-handed chiral dopant CB-15 (EM Chemicals). The helical pitch $p$ of the mixture was measured in a wedge cell and ranged from 2.33 to 60 µm(*33*). For the LC mixture with reduced bend elasticity, 4',4''-(heptane-1,7-diyl)-dibiphenyl-4-carbonitrile (CB7CB; from SYNTHON Chemicals, Germany) was additionally added to the mixture at 40 wt%(*34*).

LC cells were assembled from substrates of glass slides or coverslips coated with indium tin oxide (ITO) and treated with polyimide PI2555 and SE5661 (both from Nissan Chemicals) for planar and homeotropic boundary conditions (anchoring) for LC director, respectively. The treatment involves spin-coating polyimide onto ITO glasses, followed by baking at 180°C for 1h. PI2555-coated glass slides or coverslips were additionally rubbed to impose unidirectional planar anchoring. Silica microbeads or microcylinders were used as spacers to assemble treated ITO glasses into cells with well-defined thicknesses ranging from 7 to 60 µm by UV-activated glue. Finally, metal wires were soldered to ITO glasses as electrodes. For LC cells where an in-plane electric field parallel to the confining substrates was applied, glasses without ITO coating were polyimide-treated and used as confining substates of LCs, and a pair of ITO glasses with a gap of ~3 mm were assembled perpendicular to the substrates and soldered with wires. LC mixtures were introduced into cells via capillary force.

Electric control of LCs was achieved by connecting the electrodes of LC cells to a function generator (DS345; Stanford Research Systems) operating at a1 kHz carrier frequency to preclude complex hydrodynamic effects. To achieve an ~1 V/µm electric field in the in-plane direction, a high-voltage amplifier (Model 10/10B, Trek) was used to raise the voltage to ~3000 V. Magnetic control of LC ferromagnets was achieved by permanent neodymium magnets (K&J Magnetics) or a previously reported home-built electromagnet system containing solenoids with machined cast-iron cores driven by power supplies (BOP20-5M, Kepco)(*35*, *36*). The solenoids were arranged in Helmholtz coils to produce uniform magnetic fields up to ~30 mT. Magnetic fields used to control the dynamics of skyrmions and monopoles in LC ferromagnets were typically ~10 mT.

**Laser generation and imaging of skyrmions and monopoles**

Skyrmions and monopoles were controllably generated by laser tweezers that locally melt and align LC directors. The tweezers setup is based on an ytterbium-doped fiber laser (YLR-10-1064, IPG Photonics) and a phase-only spatial light modulator (P512-1064, Boulder Nonlinear Systems) integrated with an inverted optical microscope (IX81, Olympus)(*15*). Upon focusing the 1,064 nm infrared laser into the LC sample, local heating and optical realignment created initial $n(r)$ that eventually relaxed into skyrmions and monopoles. In a planar-alignment cell, at $d = p = 10$ μm and using a 5CB-based chiral LC mixture, in-plane skyrmions in a helical background can be reliably generated at $U \approx 1.27$ V upon laser-induced melting, while in-plane skyrmions in a uniform background can be generated at an in-plane electric field of ~1 V/μm. In a homeotropic-alignment cell, vertical torons in a uniform background can be generated without the applied electric field.

Bright-field microscopy and polarizing optical microscopy were performed using an Olympus IX-81 inverted microscope and a charge-coupled device camera (Flea-COL, from PointGrey Research)(*15*). Video-microscopy was conducted at a frame rate of 15 fps. Image analysis of monopole separation velocity was performed using the ridge detection plugin in ImageJ (National Institute of Health). Under different applied voltages, the skyrmion and monopoles as its end points in each POM video were identified. The separation velocity of the monopole-antimonopole pair at different voltages, which varied only modestly with the length of the skyrmion, was extracted as the average velocity.

Nonlinear optical imaging via three-photon excitation fluorescence polarizing microscopy (3PEF-PM) was performed using a previously reported setup to unambiguously characterize 3D $\boldsymbol{n}(\boldsymbol{r})$ configurations in LCs(*15*, *20*). Briefly, 3PEF-PM was performed using a setup built around the same IX-81 microscope with a Ti:sapphire laser (Chameleon Ultra II; Coherent) operating at 900 nm wavelength, 140 fs pulse duration, and 80 MHz repetition rate. 5CB and CB-15 molecules in the LC mixture were excited by the ultrashort pulses and the fluorescence was epi-collected by a 100✕ oil-immersion objective (NA = 1.44) and detected by a photomultiplier tube (H5784-20, Hamamatsu) after a 417/60-nm bandpass filter. The polarization state of the excitation was controlled by a pair of half-wave and quarter-wave retardation plates. In 3PEF-PM imaging, the image intensity scales as $\cos^6 \beta$, where $\beta$ is the angle between the dipole moment of the LC molecule, orientating along $\boldsymbol{n}(\boldsymbol{r})$, and the polarization of the excitation light(*37*). LC mixtures used in samples for 3PEF-PM imaging comprise additionally 15% of diacrylate nematic reactive mesogen RM 257 (Merck) and 1% of UV-sensitive photoinitiator Irgacure 369 (Sigma-Aldrich), making them partially polymerizable. Before imaging, samples were polymerized by UV illumination and the unpolymerized components were replaced with immersion oil. This process preserves the $\boldsymbol{n}(\boldsymbol{r})$ configuration of the sample, while reducing the material's birefringence by an order of magnitude, thus minimizing imaging artifacts(*38*).

**Synthesis and dispersion of magnetic nanoplates in LCs**

Barium hexaferrite $BaFe_{11}CrO_{19}$ ferromagnetic nanoplates were synthesized by the hydrothermal method described in(*39*). Briefly, 0.01 M of $Ba(NO_3)_2$, 0.04 M of $Fe(NO_3)_3 \cdot 9H_2O$, and 0.01 M of $Cr(NO_3)_3 \cdot 9H_2O$ (all from Alfa Aesar), according to the nominal stoichiometry, were dissolved in deionized (DI) water and co-precipitated by 2.72 M of NaOH (Alfa Aesar) aqueous solution with a final volume of 20 mL in a 25 mL Teflon-lined autoclave. The resulting material was hydrothermally heated to 220°C at a rate of 3°C/min, held at 220 °C for 1 h, and then cooled down to room temperature. The precipitated nanoplates were then washed with 10 wt% nitric acid and acetone and redispersed in 1 mL of DI water. The thickness of the as-synthesized nanoplates was determined to be ~10 nm and the average diameter ~105 nm using transmission electron microscopy. The nanoplates are magnetically monodomain with an magnetic moment approximately $1 \times 10^{-17}$ Am$^2$, orthogonal to their large-area faces(*10*). The magnetic nanoplates were then surface-functionalized by 5 kDa silane-terminated polyethylene glycol (JemKem Technology) to increase their stability in LCs and introduce strong homeotropic anchoring to LC molecules.

To obtain LC ferromagnets, 15 µL of LC mixtures were mixed with 15 µL of ethanol and 15 µL of 1 wt% magnetic nanoplates dispersed in ethanol. The mixture was kept at 90°C for 3 h for the ethanol to fully evaporate before being cooled rapidly to nematic phase while vigorously disturbed to reduce formation of aggregates at domain boundaries. The ensuing mixture was then centrifuged at 2,200 rpm for 5 min to remove any residual aggregates. To achieve monodomain LC ferromagnets, an unwinding electric field (~0.5 V/µm) and a colinear magnetic field perpendicular to the substrates (~15 mT) were applied while filling the LC cell(*40*). The samples were confirmed to be magnetically monodomain by the response of $\boldsymbol{n}(\boldsymbol{r})$ to externally applied magnetic fields(*35*).

**Numerical modeling**

Numerical modeling of configurations and energetics of skyrmions and monopoles was based on energy minimization of Frank-Oseen free energy that describes energetic penalties of different elastic deformations modes in $\mathbf{n}(\mathbf{r})$, supplemented by dielectric and magnetic (for LC ferromagnets) coupling terms(*9, 14, 20*). For a chiral LC, the elastic and dielectric terms read

$$F_{\text{elastic}} = \int d^3r \left\{ \frac{K_{11}}{2} (\nabla \cdot \mathbf{n}(\mathbf{r}))^2 + \frac{K_{22}}{2} [\mathbf{n}(\mathbf{r}) \cdot (\nabla \times \mathbf{n}(\mathbf{r}))]^2 + \frac{K_{33}}{2} [\mathbf{n}(\mathbf{r}) \times (\nabla \times \mathbf{n}(\mathbf{r}))]^2 \right.$$

$$\left. + \frac{2\pi K_{22}}{p} \mathbf{n}(\mathbf{r}) \cdot (\nabla \times \mathbf{n}(\mathbf{r})) \right\} \quad (1),$$

$$F_{\text{dielectric}} = -\frac{\varepsilon_0 \varepsilon_a}{2} \int d^3r \left( \mathbf{E} \cdot \mathbf{n}(\mathbf{r}) \right)^2 \quad (2).$$

Here, $\mathbf{n}(\mathbf{r})$ is the LC molecular alignment field, $K_{11}$, $K_{22}$, and $K_{33}$ are the Frank elastic constants describing the energetic costs of splay, twist, and bend deformations, respectively, and $p$ is the helical pitch of the material. Surface energy and saddle-splay deformation were not included by assuming strong boundary conditions on the surfaces, consistent with conditions in experiments. In the dielectric coupling term, $\varepsilon_0$ is vacuum permittivity, $\varepsilon_a$ is the dielectric anisotropy of the LC and $\mathbf{E}$ is the applied electric field. In a chiral LC ferromagnet, the magnetization unit vector field $\mathbf{m}(\mathbf{r}) \equiv \mathbf{M}(\mathbf{r})/|\mathbf{M}(\mathbf{r})|$, where $\mathbf{M}(\mathbf{r})$ is the local magnetization, is assumed to be collinear with $\mathbf{n}(\mathbf{r})$ due to strong homeotropic surface anchoring of LC molecules to the magnetic nanoplates. Therefore, the elastic free energy and dielectric coupling of a chiral LC ferromagnet are identical to Eqs. (1) and (2), with $\mathbf{n}(\mathbf{r})$ substituted by $\mathbf{m}(\mathbf{r})$. An additional magnetic coupling term between $\mathbf{m}(\mathbf{r})$ and an applied magnetic field $\mathbf{H}$ reads

$$F_{\text{magnetic}} = -\mu_0 |\mathbf{M}(\mathbf{r})| \int d^3r \, \mathbf{H} \cdot \mathbf{m}(\mathbf{r}). \quad (3)$$

Here, $\mu_0$ is vacuum permeability and $|\mathbf{M}(\mathbf{r})| = \rho m_p$ is the magnetization of the material (approximated to be uniform) as the product of the nanoplate density $\rho$ (10 /μm³) and the average magnetic moment of a nanoplate $m_p$ ($1 \times 10^{-17}$ Am²), both adopted from experiments (*35, 36, 41*).

The free energy was iteratively minimized using an energy-minimization routine with finite difference discretization in space and forward Euler method in time implemented in MATLAB(*14, 20*). Previously relaxed configurations or those obtained from experimental 3PEF-PM images were used as initial conditions for $\boldsymbol{n(r)}$ or $\boldsymbol{m(r)}$. Briefly, $\boldsymbol{n(r)}$ or $\boldsymbol{m(r)}$ was updated iteratively from an initial configuration using Euler-Lagrange equation until the change in the spatial average of functional derivatives converged at an energy minimum. In all simulations, the computational volume was sampled by isotropic voxels on a cubic grid at 24 gird points per $p$ and periodic boundary conditions were applied at the lateral faces of the volume.

Material parameters in simulations were chosen to match those of 5CB (Table S1) (*20*). To model the elasticity of the bend-reduced LC mixture containing 5CB and CB7CB, material parameters were set to be the same as those of 5CB, except that $K_{33}$ was reduced to a smaller value such that $K_{33}/K_{22} \leq 1$. In the limit where $K \equiv K_{11} = K_{22} = K_{33}$, Eq. (1) can be rewritten into the following form,

$$F_{\text{elastic,1-const}} = \int d^3\boldsymbol{r} \left\{ \frac{K}{2}(\nabla \boldsymbol{n(r)})^2 + \frac{2\pi K}{p}\boldsymbol{n(r)} \cdot (\nabla \times \boldsymbol{n(r)}) \right\}. \quad (4)$$

Equation (4), or Frank-Oseen free energy with the so-called one-constant approximation, has been frequently adopted to simplify calculations of LC energetics, and takes an identical functional form to the micromagnetic Hamiltonian density of solid-state chiral magnets (with the nonlocal dipole-dipole interactions neglected)(*42–44*). The connection between the continuum energy functionals of LCs and solid-state magnets suggests that similar configurations and phenomena can be anticipated in these distinct physical systems.

# Table

**Table S1.** Material parameters of 5CB.

| $K_{11}$ (pN) | $K_{22}$ (pN) | $K_{33}$ (pN) | $K$ (pN) | $\varepsilon_a$ |
|---|---|---|---|---|
| 6.4 | 3.0 | 10.0 | 6.5 | 13.8 |

# Legend for movie S1

S1. Movies showing the dynamics of monopoles at the ends of an in-plane skyrmion in a helical background at different voltages.

**Data availability:** All data generated or analyzed during this study are included in the published article and its Supplementary Information and are available from the corresponding author on a reasonable request.

**Code availability:** The codes used for the numerical calculations are available upon request.

**Acknowledgements:** We acknowledge discussions and technical assistance of Qingkun Liu and Haridas Mundoor. We are grateful to Patrick Davidson for providing CB7CB used at the initial stages of this study.

**Funding:** This research was supported by the U.S. National Science Foundation grant DMR-1810513.

**Author Contributions:** J.-S.B.T. performed experiments. A.J.H. synthesized magnetic nanoplates. J.-S.B. T. and J.-S.W. performed numerical modelling. I.I.S. directed research and provided funding. All authors contributed to the writing of the manuscript.

**Competing interests:** The authors have no competing interests.